\documentclass[sigconf]{acmart}

\AtBeginDocument{%
  \providecommand\BibTeX{{%
    \normalfont B\kern-0.5em{\scshape i\kern-0.25em b}\kern-0.8em\TeX}}}

\copyrightyear{2023} 
\acmYear{2023} 
\setcopyright{acmlicensed}\acmConference[RecSys '23]{Seventeenth ACM Conference on Recommender Systems}{September 18--22, 2023}{Singapore, Singapore}
\acmBooktitle{Seventeenth ACM Conference on Recommender Systems (RecSys '23), September 18--22, 2023, Singapore, Singapore}
\acmPrice{15.00}
\acmDOI{10.1145/3604915.3608766}
\acmISBN{979-8-4007-0241-9/23/09}

\usepackage{algorithm}
\usepackage{bm}
\usepackage{algorithmicx}
\usepackage{algpseudocode}
\usepackage{subcaption}
\usepackage{booktabs}
\usepackage{amsmath, amsthm}
\usepackage{multirow}

\theoremstyle{definition}

\usepackage{courier}  
\usepackage{natbib}  
\usepackage{caption} 
\usepackage{listings}
\usepackage{makecell}
\lstdefinestyle{mystyle}{
    basicstyle=\tiny,
    commentstyle=\color{codegreen},
    keywordstyle=\color{magenta},
    numberstyle=\tiny\color{codegray},
    stringstyle=\color{codepurple},
    basicstyle=\fontsize{8.5}{9}\selectfont\ttfamily,
    breakatwhitespace=false,         
    breaklines=true,                 
    captionpos=b,                    
    keepspaces=true,                 
    numbers=left,                    
    numbersep=5pt,                  
    showspaces=false,                
    showstringspaces=false,
    frame = single
}

\usepackage{pifont}
\usepackage{color}
\newcommand{\cmark}{\ding{51}}%
\newcommand{\xmark}{\ding{55}}%
\acmSubmissionID{35}
\begin{document}

\title{Rethinking Multi-Interest Learning for Candidate Matching in Recommender Systems}


\author{Yueqi Xie}
\affiliation{%
  \institution{HKUST}
  \country{Hong Kong}}
\email{yxieay@connect.ust.hk}

\author{Jingqi Gao}
\affiliation{%
  \institution{Upstage}
  \country{Hong Kong}
}
\email{mrgao.ary@gmail.com}

\author{Peilin Zhou}
\affiliation{%
  \institution{HKUST(gz)}
  \country{China}}
  \email{zhoupalin@gmail.com}

\author{Qichen Ye}
\affiliation{%
  \institution{Peking University}
  \country{China}}
  \email{yeeeqichen@pku.edu.cn}

\author{Yining Hua}
\affiliation{%
  \institution{MIT}
  \country{USA}
}
\email{ninghua@mit.edu}

\author{Jae Boum Kim}
\affiliation{%
  \institution{HKUST}
  \country{Hong Kong}}
  \email{jbkim@cse.ust.hk}

\author{Fangzhao Wu}
\affiliation{%
  \institution{MSRA}
  \country{China}}
  \email{wufangzhao@gmail.com}
  
\author{Sunghun Kim}
\affiliation{%
  \institution{HKUST}
  \country{Hong Kong}}
  \email{hunkim@cse.ust.hk}
  
\renewcommand{\shortauthors}{Y. Xie et al.}
\renewcommand{\shorttitle}{Rethinking Multi-Interest Learning for Candidate Matching}
\begin{abstract}
Existing research efforts for multi-interest candidate matching in recommender systems mainly focus on improving model architecture or incorporating additional information, neglecting the importance of training schemes. 
This work revisits the training framework and uncovers two major problems hindering the expressiveness of learned multi-interest representations.
First, the current training objective (i.e., uniformly sampled softmax) fails to effectively train discriminative representations in a multi-interest learning scenario due to the severe increase in easy negative samples. 
Second, a \textit{routing collapse} problem is observed where each learned interest may collapse to express information only from a single item, resulting in information loss. 
To address these issues, we propose the REMI framework, consisting of an \textbf{I}nterest-aware \textbf{H}ard \textbf{N}egative mining strategy (IHN) and a \textbf{R}outing \textbf{R}egularization (RR) method. 
IHN emphasizes interest-aware hard negatives by proposing an ideal sampling distribution and developing a Monte-Carlo strategy for efficient approximation. RR prevents \textit{routing collapse} by introducing a novel regularization term on the item-to-interest routing matrices.
These two components enhance the learned multi-interest representations from both the optimization objective and the composition information.
REMI is a general framework that can be readily applied to various existing multi-interest candidate matching methods. 
Experiments on three real-world datasets show our method can significantly improve state-of-the-art methods with easy implementation and negligible computational overhead. 
The source code is available at \url{https://github.com/Tokkiu/REMI}.
\end{abstract}




\maketitle

\section{Introduction}
\label{sec-intro}
Recommender Systems (RS) play a crucial role in various online services~\cite{covington2016deep, guy2010social}, such as E-Commerce platforms~\cite{hwangbo2018recommendation,li2019multi,xie2022decouple,zhou2022equivariant,zhang2023AttenMixer}. 
They are typically comprised of two stages: candidate \textit{matching} and \textit{ranking}~\cite{chai2022user,cen2020controllable}.
The candidate matching stage aims to efficiently retrieve thousands of items from the large item corpus via accurately modeling user interests, laying the foundation for fine ranking~\cite{gomez2015netflix,chai2022user,lv2019sdm}. 
Due to its fundamental importance and specific requirements such as high efficiency, candidate matching has attracted increasing research interest~\cite{covington2016deep,li2021path}.


\begin{figure}[t]
    \begin{subfigure}{\linewidth}
        \centering
\includegraphics[width=1.0\linewidth]{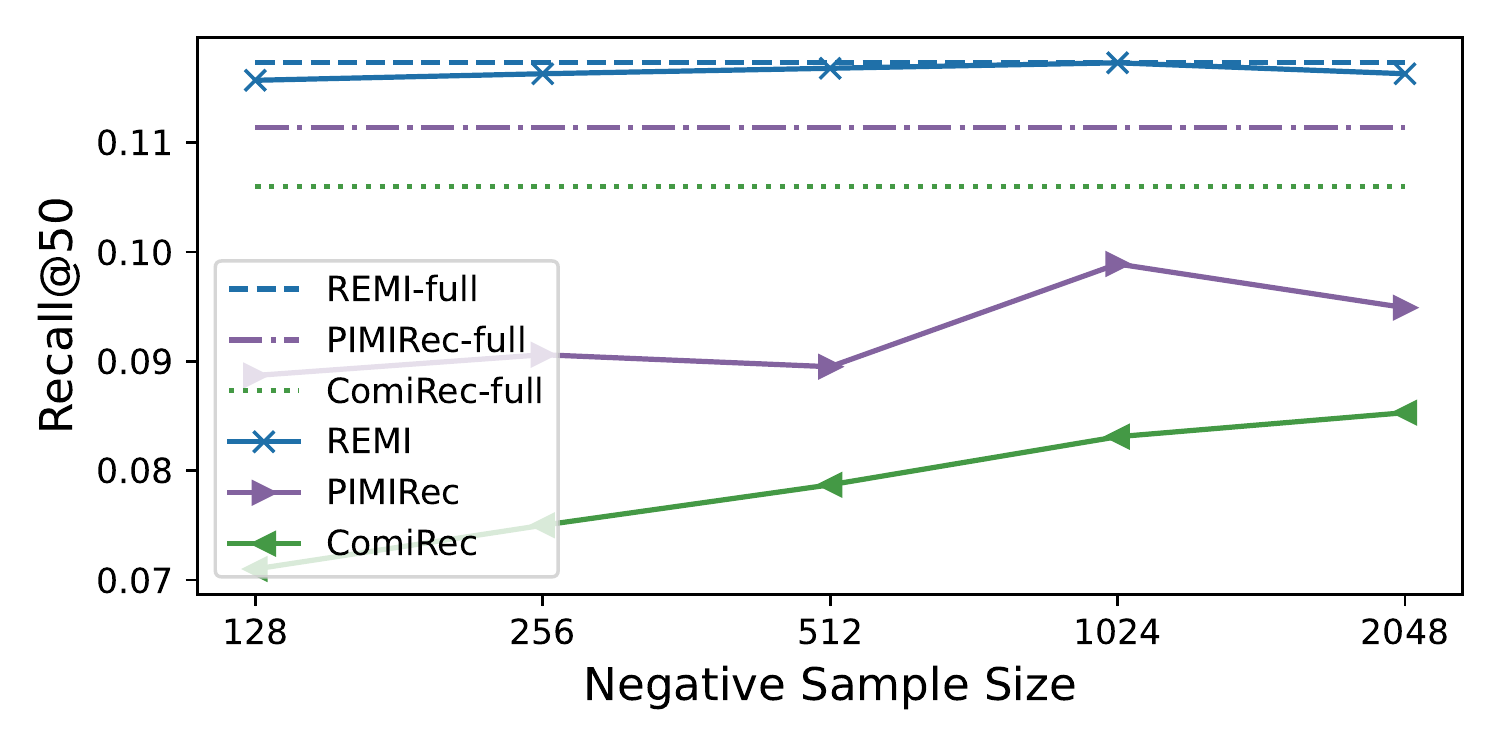}
\subcaption{}
    \end{subfigure}
    \centering
        \begin{subfigure}{\linewidth}
        \centering
\includegraphics[width=1.0\linewidth]
{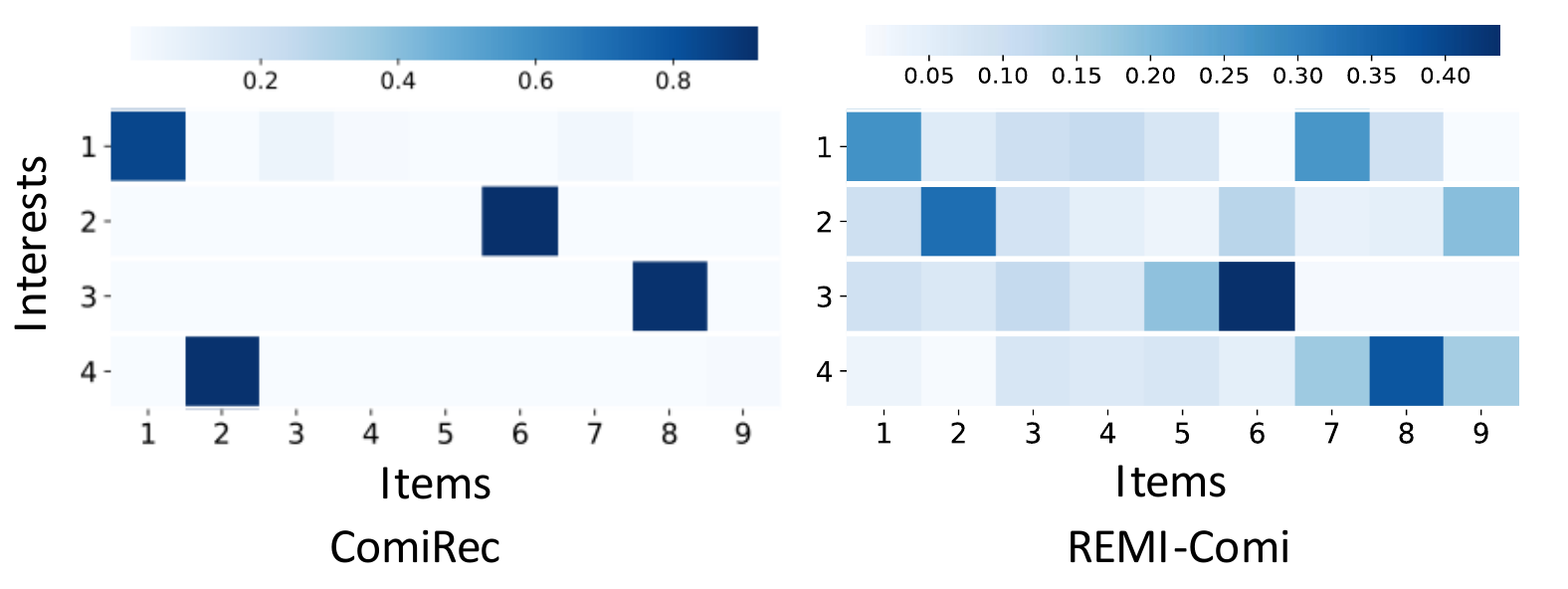}
\subcaption{}
    \end{subfigure}
    \caption{(a): Performance of different methods with practical negative sampling sizes, compared with full softmax (shown as the dashed lines, impractical in real-world scenarios) on Amazon Book. Multi-interest models, i.e., ComiRec and PIMIRec, suffer from dramatical degradation when using uniformly sampled softmax. Training with our REMI framework alleviates the problem and further improves performance. (b): 
    Visualization of an example of item-to-interest routing weights of ComiRec and REMI-enhanced version. REMI helps avoid \textit{routing collapse}.}    
    \label{fig:intro_sample}
\end{figure}

Recently, multi-interest learning-based methods~\cite{li2019multi,cen2020controllable} have shown great potential in improving matching performance.
These methods explicitly generate users' diverse interest representations from their behavior sequences, breaking the representation bottleneck of using a single generic user embedding.
MIND~\cite{li2019multi} first captures the user's multiple interests through dynamic routing with Capsule Network~\cite{sabour2017dynamic}.
Afterward, ComiRec~\cite{cen2020controllable} takes diversity into consideration  and additionally leverages multi-head attention to encode users' diverse interests.
Several recent works~\cite{Chen2021PIMI, chai2022user} have further enabled awareness of periodicity, interactivity, and user profile.

\begin{figure*}[t]
  \centering
  \includegraphics[width=0.97\textwidth]{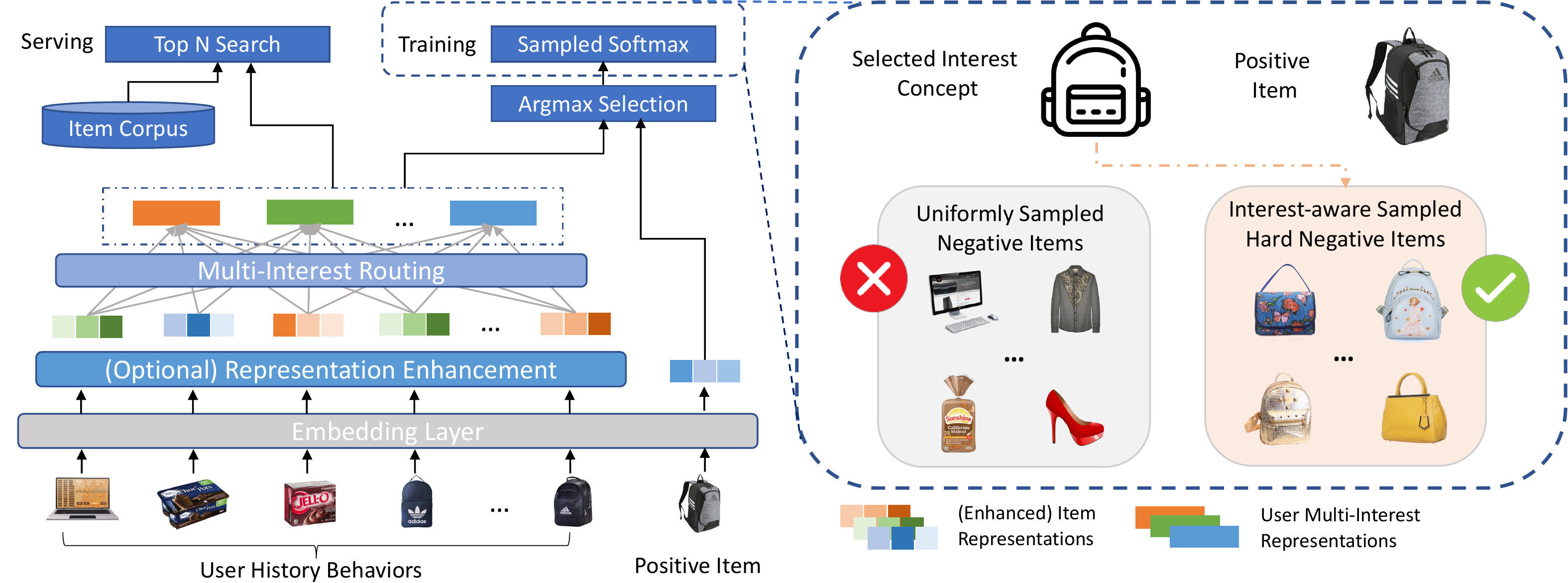}
  \caption{Demonstration of the existing framework (Left) and illustration of different sampling strategies (Right). Although some items in the uniformly sampled negative items (e.g., bread, computer) can be informative hard negative samples when training a general user representation for the sequence consisting of food, electronic devices, and bags, they become unrelated easy negative samples when training the selected interest representation related to ``bag'' in the existing multi-interest training scheme. The interest-aware hard negative sampling (IHN) alleviates this problem by prioritizing hard negative samples related to the current-selected interest with an adjustable degree, which benefits the learning of discriminative user interest representation. 
  }
\label{fig:overview}
\end{figure*}

Despite various model architectures and information explored in multi-interest learning, few efforts are devoted to the training scheme. 
Previous works generally follow a similar scheme, as depicted in Figure~\ref{fig:overview} (Left).
The user behavior sequence is first encoded into item embeddings and then routed to multiple interest concept representations. 
To train these multi-interest representations, a common practice is to use the positive target item for each user to select the interest representation closest to the true label.
This interest representation is then activated as the user representation and trained similarly to general candidate matching, where a uniformly or log-uniformly sampled softmax~\cite{covington2016deep,jean2014using} is applied to efficiently train the model on a large corpus.

%
In this work, we \textbf{RE}visit the currently used \textbf{M}ulti-\textbf{I}nterest learning framework and uncover two major problems. 
First, we raise doubts about the effectiveness of uniformly sampled softmax in multi-interest scenarios.
While uniformly sampled softmax has been shown to be effective in training general recommendation systems~\cite{wu2022effectiveness}, it falls short in multi-interest recommendation systems, such as ComiRec~\cite{cen2020controllable} and PIMIRec~\cite{Chen2021PIMI}. As illustrated in Figure~\ref{fig:intro_sample} (a),  it has significantly worse performance compared to full softmax within a reasonable sample size range (e.g., below a thousand).
This is due to  the \textit{selected-interest-focused training scheme} in multi-interest learning that the training focuses on a specific representation of one of the user's multiple interests instead of a general representation of the user.
A specific interest concept representation is more susceptible to encountering "easy negatives" compared to an overall preference representation. 
For example, as shown in Figure~\ref{fig:overview}, consider a user whose historical sequence includes food, electronics, and bags, and the concept related to ``bag'' is activated for training based on the target item.
If we apply uniform sampling, most negatives are easily distinguishable from the representation of the ``bag'' concept and are therefore ``easy negatives'',
even though some of them may be informative with respect to the overall preference of the user.
A lack of informative hard negatives has been proven to impact training~\cite{grbovic2018real,chen2022generating}.
Therefore, uniformly sampled softmax is incapable of training expressive multi-interest representations.



Second, multi-interest models rely on item-to-interest routing to determine the items that form an interest. This routing module is usually a multi-head attention~\cite{cen2020controllable} or capsule network~\cite{li2019multi}. Although the multi-head attention-based routing strategy exhibits higher efficiency and comparable effectiveness than capsule networks~\cite{cen2020controllable,Chen2021PIMI},  we observe that it suffers from \textit{routing collapse}. 
This means that each interest representation may only be made up of one most relevant item instead of the intended all relevant items.
For instance, the representation of interest in ``bags'' would consist of only one specific bag, not all related bags in the user sequence.
Figure~\ref{fig:intro_sample} (b) shows an example visualization of normalized routing logits with ComiRec-SA~\cite{cen2020controllable}. 
This can result in information loss in the interest representation from the beginning of its formation, hindering the ability to learn expressive multi-interest representations.

To solve these issues, we propose the REMI framework with Interest-aware Hard Negative mining (IHN) and Routing Regularization (RR).
First, to alleviate the easy negative problem, we design an ideal distribution that prioritizes, to an adjustable extent, the hard negative items with higher scores related to the currently chosen interest concept (Figure~\ref{fig:overview}, Right).
However, direct sampling from this distribution leads to high complexity in computing all the relevance scores and per-user sampling. Therefore, we further propose to leverage the Monte-Carlo importance sampling technique~\cite{robert1999monte,kloek1978bayesian} to approximate the above ideal distribution with a few lines of extra code and negligible computation overhead.
For the \textit{Routing Collapse} problem, we offer a simple and effective routing regularization strategy on the variance of item-to-interest routing weights for each interest. 
The two modules improve the expressiveness of the multi-interest representation from two aspects: the optimization objective of training and the composition of information.
As demonstrated in Figure~\ref{fig:intro_sample}, the proposed solution not only enable the effectiveness of sampled softmax within a reasonable sampling size but also further improve the achievable performance. Also, the visualization verifies the effectiveness of REMI in solving the \textit{routing collapse} problem.


We conduct extensive experiments on three real-world, large-scale public recommendation datasets. 
The results show that REMI significantly boosts the performance of existing multi-interest candidate matching methods when applied to them.
In addition, in-depth analyses of two main components of our framework demonstrate their effectiveness in solving the targeted problems. 
Our contributions are summarized as follows:
\begin{itemize}
\item We revisit the training scheme of multi-interest learning and uncover two major problems regarding the current negative sampling strategy and routing collapse, which significantly hinder the performance of multi-interest models.
    \item We propose a simple and effective interest-aware hard negative mining strategy and routing regularization to solve the identified issues and improve the expressiveness of the multi-interest representation from two aspects: the optimization objective of training and the composition of information.
    \item We conduct extensive experiments which demonstrate that REMI significantly improves the existing multi-interest candidate matching solutions with a few lines of extra code and negligible computation overhead. 
\end{itemize}

\section{Related Work}
\subsection{Candidate Matching}
Candidate matching is an essential step in large-scale industrial RS, as it efficiently filters a subset from a large item pool for subsequent fine ranking~\cite{gomez2015netflix,chai2022user,lv2019sdm}.
Due to its high demand for efficiency, candidate matching models typically utilize lightweight architecture and avoid candidate awareness during user modeling.
In the early stage, Collaborative Filtering (CF) based solutions~\cite{sarwar2001item,kabbur2013fism} introduce learnable matching between user and candidates and Neural Collaborative Filtering (NCF)~\cite{NCF} enhances the classical CF with multi-layer perceptrons.
Afterward, the two-tower DNN structures~\cite{covington2016deep,web_search}  become highly popular with high computation efficiency, which avoids early interaction between user modeling and candidate modeling.
In addition, tree-based structures and graph-based structures have also been explored for deep candidate matching~\cite{tree-based1,tree-based2,PDNP}. For example, PDNP~\cite{PDNP} builds a retrieval architecture based on a 2-hop graph, which enables online retrieval with low latency and computation cost.
These solutions generally model the user’s preference in one vector, potentially limiting the representation ability considering the multi-interest nature of users.

\subsection{Multi-Interest Learning}
Recent research has found that modeling users' interests as a single vector may be insufficient to accurately capture the complex interaction patterns of users~\cite{li2019multi,cen2020controllable}.
As a result, multi-interest learning is gradually gaining more attention in  both the matching stage and ranking stage of RS.
Targeting the matching stage, MIND~\cite{li2019multi} first adopts a dynamic routing mechanism to aggregate users' historical behaviors into multiple interest capsules~\cite{sabour2017dynamic}.
Afterward, ComiRec~\cite{cen2020controllable} further investigates multi-head attention-based multi-interest routing for capturing the user's diverse interests and introduces diversity controllable methods.
Under a similar framework, PIMIRec~\cite{Chen2021PIMI} and UMI~\cite{chai2022user} incorporate time information, interactivity, and user profile. Re4~\cite{zhang2022re4} further takes the backward flow into account to regularize the process.  Despite improved architectures and enriched information, these works generally follow the training of the general training scheme of candidate matching, where we identify several problems in our work. 
For the ranking stage, DMIN~\cite{dmin}, DemiNet~\cite{deminet}, and MGNM~\cite{tian2022multi} explore more architectures and enable candidate-aware Click-through Rate Prediction through a multi-head attention mechanism. Note that in the experimental section, DMIN, DemiNet, and MGNM are not chosen as baselines due to the inapplicable computational complexity for the matching stage.
\subsection{Negative Sampling in Recommender Systems}
Negative sampling is crucial in RS because it enables the selection of a small set of negative examples instead of using all the non-target items. This reduces the computational cost and speeds up training while still enabling the model to learn meaningful representations.
In the early stages, Bayesian Personalized Ranking~\cite{rendle2012bpr} uniformly samples negative items from items that users had not interacted with. Later, several studies~\cite{togashi2021alleviating,he2016fast} give popular items higher probabilities of being sampled as negative items to suppress popularity bias.
However, popularity-based solutions cannot effectively help solve the problem of increasing easy negative samples in multi-interest learning.
MNS~\cite{yang2020mixed} mixes batch and uniformly sampled negatives to mitigate the selection bias but not targets hard negative samples.
Some enhanced methods~\cite{rafailidis2019bayesian,liu2019geo,ding2019reinforced,zheng2021dgcn} leverage additional information to improve negative sampling. 
Specifically, these methods produce informative negative samples for RS in certain application scenarios with available information such as connections in social networks~\cite{rafailidis2019bayesian,wang2016social,yu2021socially,zhao2014leveraging}, user locations~\cite{liu2019geo,manotumruksa2017personalised,yuan2016joint}, item category information~\cite{zheng2021dgcn}, and detailed user feedback~\cite{loni2016bayesian,ding2019reinforced,ding2018improved}.
However, these solutions are only applicable in certain scenarios and cannot produce hard negative samples at scale based on current user interest representations.
Finally, the most related hard negative sampling strategies are those that produce negative samples according to the current user representation~\cite{zhang2013optimizing,rendle2014improving,chen2022learning,chen2022generating,ding2020simplify}. 
For example, the work~\cite{zhang2013optimizing} uses a dynamic rejection sampling strategy according to item ranking, the work~\cite{ding2020simplify} leverages score-based memory update and variance-based sampling, and a recent work~\cite{chen2022generating} generates hard negative samples for sequential recommendation using the Next Negative Item (NNI) Sampler with Pre-selection and Post-selection.
Generally, these solutions allow dynamic sampling by developing complex schemes to explicitly perform context-aware sampling. Our method, in contrast, simply re-weights the loss generated with uniform sampling and effectively addresses the problem in multi-interest learning with negligible cost.

\section{Preliminary}
\subsection{Problem Formulation}
Suppose we have a set of users denoted as $\mathcal{U}$ and a large item corpus denoted as $\mathcal{I}$. For each user $u \in \mathcal{U}$, we have a historical behavior sequence $\mathbf{s^{(u)}}=(i^{(u)}_{1}, i^{(u)}_{2}, \cdots, i^{(u)}_{n})$ sorted by time, where $n$ is the maximum sequence length and $i^{(u)}_{t}$ represents the $t$-th item in the user behavior sequence.
The candidate matching stage in RS aims to 
efficiently retrieve a subset of items the user is likely to interact with from the huge item corpus $\mathcal{I}$. 

\subsection{Existing Framework}
\subsubsection{Model Architecture}
Current solutions~\cite{li2019multi,cen2020controllable} provide several strategies to form the multi-interest extraction network to get the representation matrices of the multiple interests. Since we focus on the training scheme, we simply abstract the process with two steps 1) item representation (including item embedding and optional representation enhancement) and 2) multi-interest routing (Figure~\ref{fig:overview}). 

Let $d$ denote the hidden dimension of the representation, $n$ denote the max sequence length, and $K$ denote the total number of interests, we write step 1) as: 
\begin{equation}
\mathbf{H} = \mathcal{F}(s^{(u)}),
\end{equation}
where $\mathcal{F}$ denotes the model function to encode (and enhance) the items in the user behavior sequence, and $\mathbf{H} \in \mathbb{R}^{d \times n}$ denotes the encoded (and enhanced) item representations. And the learned item embedding matrix for the items in the corpus is denoted as $\mathbf{E}=[\mathbf{e}_1, \mathbf{e}_2,\cdots,\mathbf{e}_{|\mathcal{I}|}] \in \mathbb{R}^{|\mathcal{I}| \times d}$.

Then, step 2) can be formulated as calculating the item-to-interest routing matrix $\mathbf{A}\in\mathbb{R}^{n\times K}$ followed by obtaining the final multi-interest representations $\mathbf{V}_u\in\mathbb{R}^{d\times K}$:
    \begin{align}
        &\mathbf{A} = \mathcal{G}(\mathbf{H}),
        \\
    &\mathbf{V}_u = \mathbf{H}\mathbf{A},
    \end{align}
where $\mathcal{G}$ is the function to get the routing matrix from the (enhanced) item representations. 

\subsubsection{Training Scheme}
During training, an \textit{argmax} operator is then used to select the most relevant user embedding vector $\mathbf{v}_u$ according to the positive item $i^+$:
\begin{equation}
    \label{eqn:argmax}
    \begin{aligned}
        & k= \operatorname{argmax}\left(\mathbf{V}_u^{\top} \mathbf{e}_{i^+}\right),
        \\
        &\mathbf{v}_u = \mathbf{V}_u[:, k],
    \end{aligned}
\end{equation}
where $\mathbf{e}_{i^+}$ is the embedding of the target item $i^+$, and $\mathbf{v}_u\in \mathbb{R}^{d \times 1}$ is the selected interest concept representation.
Then, the problem can be modeled as a classification task. The likelihood of the user $u$, with the chosen interest $k$, 
interacting with the target (positive) item $i^{+}$ can be represented with a softmax function:
\begin{equation}
\begin{split}
\label{eqn:likelihood}
    P(i^+|u) = \frac{e^{\mathbf{v}_u^\top \mathbf{e}_{i^+}}}{\sum_{i\in\mathcal{I}}e^{\mathbf{v}_u^\top \mathbf{e}_i}}
=\frac{e^{\mathbf{v}_u^\top \mathbf{e}_{i^+}}}{e^{\mathbf{v}_u^\top \mathbf{e}_{i^+}} + \sum_{i^-\in\mathcal{I} \setminus {\{i^+\}}}e^{\mathbf{v}_u^\top \mathbf{e}_{i^-}}}.
\end{split}
\end{equation}

The loss function of the model is to minimize the negative log-likelihood of all the positive pairs $(u, i^+)$ in the training dataset $\mathcal{D}$:
\begin{equation}
    \label{eqn:loss}
    \mathcal{L_{SM}(\theta)} =\frac{1}{\mid \mathcal{D} \mid} \sum_{(u, i^+)\in \mathcal{D}} -log P(i^+|u).
\end{equation}

Due to the computation complexity of the denominator with huge amounts of negative items  in \autoref{eqn:likelihood}, multi-interest learning typically follows the common practice in general large-corpus candidate matching tasks to use a generic sampled softmax technique~\cite{jean2014using, covington2016deep} with log-uniform sampling or uniform sampling to train the model. We detailly discuss the sampled softmax strategy in Section~\ref{sampled-sm}.

\subsubsection{Serving Scheme}
\label{pre-serve}
For serving, the multi-interest representations $\mathbf{V}_u$ are obtained for each user $u$. Then each interest embedding retrieves top-N items from $\mathcal{I}$ based on the inner production using Faiss~\cite{johnson2019billion}. Afterward, the final score $f(u,i)$ for the item $i$ in the retrieved set for user $u$ is calculated as:
\begin{equation}
    f(u, i)=\max _{1 \leq k \leq K}({v_u^{k}}^{\top} \mathbf{e}_i ),
\end{equation}
where $v_u^{k}\in \mathbb{R}^{d \times 1}$ is the $k$-th interest representation. Finally, the top-N scored items are recommended to the user $u$.
\subsection{Sampled Softmax Loss}
\label{sampled-sm}
This section revisits the sampled softmax strategy in a multi-interest learning context. 
Sampled softmax function reduces computation complexity by only considering the positive class and sampled negative classes. 
A smaller negative sample size indicates higher training efficiency.
A practical sampling size would be within thousands. Note that in prior works~\cite{cen2020controllable, Chen2021PIMI}, the negative sampling size is claimed to be 10 in the paper, but their official codes\footnote{https://github.com/THUDM/ComiRec} \footnote{https://github.com/ChenGaoDe/PIMI\_Rec} show that the sample size is set to $10 * batch size$ per batch, and the negative samples are shared in the batch. Therefore, the negative sample size is commonly set to $1280$ to pair with one positive training sample in practice in the prior studies.

There are multiple variants of sampled softmax~\cite{jean2014using, covington2016deep}, and the prior studies utilize the version implemented in TensorFlow~\cite{tensorflow2015-whitepaper}:
\begin{equation}
\label{equ:sampled_softmax}
\begin{split}
     \mathcal{L_{SSM}(\theta)} = \frac{1}{|\mathcal{D}|} \sum_{(u, i^+) \in \mathcal{D}} 
    -\log\frac{e^{\mathbf{v}_u^\top \mathbf{e}_{i^+}-\log Q(i^+)}}
    {e^{ \mathbf{v}_u^\top \mathbf{e}_{i^+}-\log Q(i^+)}+\sum_{i=1}^{L} e^{ \mathbf{v}_u^\top \mathbf{e}_{i^-}-\log Q(i^-)}},
\end{split}
\end{equation}
where $L$ is the negative sampling size, and $i^-$ is sampled from the distribution $Q$. $\log Q$ is the correction term in sampled softmax and can be simply removed with uniform sampling or the strategy in~\cite{jean2014using}. 
In prior works~\cite{cen2020controllable,Chen2021PIMI,chai2022user,zhang2022re4}, $Q$ is often generic sampling distributions such as log-uniform and uniform sampling, which are shown to perform relatively well in general RS learning~\cite{wu2022effectiveness}. Specifically, using log-uniform sampling on the item set sorted by popularity gives the popular items a higher probability of being selected as negative samples.  

However, in multi-interest scenarios, such a solution does not perform well due to the \textit{selected-interest-focused training scheme} in multi-interest learning.
As discussed in \autoref{sec-intro}, uniformly or log-uniformly sampled items are more likely to be easy negatives for the selected specific interest from multiple interests compared with general user representation, reducing the training effectiveness of the uniformly sampled softmax function.

\section{Methodology}
REMI is a general training framework which 
can be readily applied to various multi-interest learning methods~\cite{cen2020controllable,Chen2021PIMI,chai2022user} without architecture modification and boost their performance. 
In this section, we detailly introduce two main components of REMI, interest-aware negative mining (IHN) and routing regularization (RR), which solve the problem of the increase of easy negatives and routing collapse in multi-interest learning, respectively. Then we summarize the overall training objective.

\subsection{Interest-aware Hard Negative Mining (IHN) for Multi-Interest Learning}
\label{method-ihn}
To alleviate the problem caused by the severe increase of easy negatives in multi-interest scenarios with generic sampled softmax, we first design an ideal sampling distribution, followed by a strategy to approximate the ideal sampling with  only a few lines of extra codes and negligible computation overhead.


\textbf{Ideal Sampling Distribution Design.} 
We follow two principles in designing the ideal distribution:
\begin{itemize}
    \item \textit{The informative negative samples are those the selected interest concept is likely to be misclassified to interact with.}
    \item \textit{
    The degree of hardness should be adjustable since the optimal effective hardness of negative samples depends on the sample size and dataset. }
\end{itemize}

Therefore, we propose a negative sampling distribution $q_\beta$, which is subjective to the chosen interest concept $\mathbf{v}_u$, and assign the hard negative items with higher probabilities of being sampled. Inspired by the sampling strategy in contrastive learning~\cite{robinson2020contrastive}, we propose $q_\beta$ as follows:
\begin{equation}
    q_\beta(i^-) \propto e^{(\beta \mathbf{v}_u^\top \mathbf{e}_{i^-})},
\end{equation}
with concentration parameter $\beta\ge 0$. 
The inner product measures the similarity between the interest and item embeddings. A larger inner product indicates a greater similarity and a harder example.
$\beta$ controls the hardness level for optimal training effectiveness. 
Then the objective in the sampled softmax in \autoref{equ:sampled_softmax} can be re-written as:
\begin{equation}
    \label{equ:ssm_ours}
   \mathcal{L_{SSM}(\theta)}= \frac{1}{|\mathcal{D}|} \sum_{(u, i^+) \in \mathcal{D}}-\log 
\frac{e^{\mathbf{v}_u^\top \mathbf{e}_{i^+}}}{e^{ \mathbf{v}_u^\top \mathbf{e}_{i^+}}+L  \mathbb{E}_{i^{-} \sim q_{\beta}}\left[e^{ \mathbf{v}_u^\top \mathbf{e}_{i^-}}\right] },
\end{equation}
where the correction term is omitted for simplicity. 
However, directly sampling from $q_\beta$ also suffers from great computation complexity to calculate all the $e^{\beta \mathbf{v}_u^\top \mathbf{e}_{i^-}}$ for huge amounts of items. 
In fact, its computation complexity is equivalent to full softmax.
As a result, our next research question becomes: Is it possible to approximate $\mathbb{E}_{i^{-} \sim q_{\beta}}\left[e^{ \mathbf{v}_u^\top \mathbf{e}_{i^-}}\right] $ without explicitly sampling from $q_{\beta}$?

\textbf{Importance Sampling-based Approximation.}
To approximate the ideal distribution at a low cost, we resort to the Monte-Carlo importance sampling techniques~\cite{robert1999monte,kloek1978bayesian} to sample from a simple uniform distribution $p$ as follows:
\begin{equation}
\label{equ:approximation}
	\mathbb{E}_{i^{-} \sim q_{\beta}}\left[e^{ \mathbf{v}_u^\top \mathbf{e}_{i^-}}\right] 
=\mathbb{E}_{i^{-} \sim p}\left[e^{\mathbf{v}_u^\top \mathbf{e}_{i^-}} q_{\beta} / p\right]
=\mathbb{E}_{i^{-} \sim p}\left[e^{(\beta+1) \mathbf{v}_u^\top \mathbf{e}_{i^-}} / Z_{\beta}\right],
\end{equation}
where $Z_{\beta}$ is the partition function that can be empirically estimated over $p$:
\begin{equation}
\widehat{Z}_{\beta}=\mathbb{E}_{i^{-} \sim p}\left[e^{\beta \mathbf{v}_u^\top \mathbf{e}_{i^-}}\right]=\frac{1}{L} \sum_{i=1}^{L} e^{\beta \mathbf{v}_u^\top \mathbf{e}_{i^-}}.
\end{equation}
Since the approximation only re-weights the objective for each sample and keeps the sampling procedure a batch-wise uniform sampling, such a method only leads to a negligible computation overhead. In addition, the implementation only requires a few lines of extra code and can be readily incorporated into the existing training framework (pseudocode provided in \autoref{supp-code}).

\textbf{Analysis.}  We then discuss the influence of concentration parameter $\beta$ starting from two extreme cases. 
When $\beta = 0$, our solution is equivalent to uniform sampling since all the items have the same probability to be sampled.
When $\beta \rightarrow \infty$, the loss function tends to focus on the hardest negative samples for a certain user since the hardest samples dominate the calculation of expectation in \autoref{equ:approximation}. Note that focusing on the hardest negative samples does not necessarily lead to better effectiveness since there are false negative samples in recommendation scenarios, and easy negatives are also important for training recommendation models~\cite{gomez2015netflix}.
Choosing a proper $\beta$ 
between two extreme cases to improve the effectiveness and efficiency of sampled softmax-based training. 
We include more empirical case studies on $\beta$ in Section~\ref{exp-rq4-hyper}.

\subsection{Routing Regularization for Routing Collapse}

This section discusses the \textit{Routing Collapse} problem and provides a simple yet effective regularization-based solution.

\textbf{Routing Collapse.} 
A key module in multi-interest learning is items-to-interests routing, which is typically achieved with multi-head attention or capsule networks~\cite{cen2020controllable, li2019multi}.
Multi-head attention-based multi-interest routing demonstrates higher efficiency and comparable performance compared with capsule networks~\cite{Chen2021PIMI,cen2020controllable}, and therefore they are widely applied as the base structure in recent work~\cite{Chen2021PIMI,chai2022user,cen2020controllable,zhang2022re4}.
However, we observe that after training for multiple epochs, the interests tend to over-focus on single items in the behavior sequence, as shown in Figure~\ref{fig:intro_sample} (b).
In this case, only a small portion of the items in the user history is considered, which impacts the expressiveness of multi-interest representations from the beginning of its composition information.
The prior study~\cite{dong2021attention} on attention mechanism also observes similar problems when solely applying attention without MLP.
Such collapse can be viewed as the model falling to a locally optimal solution.

\textbf{Routing Regularization.} We then propose a simple and effective solution to avoid \textit{Routing Collapse} in our scenarios. Unlike most regularization done in the representation space~\cite{tan2021sparse,zhang2022re4}, we find that the collapse was caused by the sparsity of item-to-interest routing matrix $\mathbf{A}\in\mathbb{R}^{n\times K}$, so we introduce the variance regularizer on the routing weights to eliminate sparsity and effectively address the problem.
Specifically, we calculate the regularization term with the following equations:
\begin{equation}
\begin{split}
        &\mathbf{C}=(\mathbf{A}-\bar{\mathbf{A}})^{\top}(\mathbf{A}-\bar{\mathbf{A}}), \\
&\mathcal{L}_{reg}=\|\operatorname{diag}(\mathbf{C})\|_{F}^{2},
\end{split}
\end{equation}
where $\bar{\mathbf{A}}$ is the mean of $\mathbf{A}$ along the first axis,
$\mathbf{C}\in\mathbb{R}^{K\times K}$ represents the covariance matrices of routing weights for different interests, $\operatorname{diag}(\mathbf{C})$ represents the extraction and construction of a diagonal matrix, and $\|\cdot\|_{F}$ denotes the Frobenius norm of matrices.
Overall, our training objective can be formed as
\begin{equation}
    \mathcal{L} = \mathcal{L}_{SSM}+ \lambda\mathcal{L}_{reg},
\end{equation}
where $\lambda$ is the hyper-parameter that balances two losses. The inference procedure is the same as prior works~\cite{cen2020controllable,zhang2022re4}, as introduced in Section~\ref{pre-serve}.


\section{Experiments}
We conduct experiments on three large-scale real-world  datasets to answer the following research questions: 
\begin{itemize}
    \item \textbf{RQ1:} 
    Can REMI help achieve state-of-the-art performance on candidate matching?
    \item \textbf{RQ2:} 
    Can REMI be combined with different multi-interest learning models and boost their performance?
    \item \textbf{RQ3:} Can the Interest-aware Hard Negative mining strategy (IHN) solve the problems regarding sampled softmax for effective and efficient training? 
    \item \textbf{RQ4:} Can Routing Regularization (RR) solve the problem regarding routing collapse? 
    \item \textbf{RQ5:} What is the effect of different components and hyper-parameters in REMI?
\end{itemize}
\subsection{Experimental Settings}
\subsubsection{Dataset.}
\begin{table}
  \centering
  \caption{\label{tab:match_stats} Statistics of datasets.}
  \begin{tabular}{c|c|c|c}
    \hline \hline
    \textbf{Dataset} & \# users & \# items & \ \# interactions \\
    \hline
    Amazon Books & 603,668 & 367,982 & 8,898,041 \\
    Gowalla & 65,506 & 174,605 & 2,061,264 \\
    RetailRocket & 33,708 & 81,635 & 356,840 \\
    \hline \hline
  \end{tabular}
\end{table}

\label{exp-setup}

We select three large-scale public datasets to evaluate the effectiveness of REMI:
\begin{itemize}
    \item \textbf{Amazon}~\cite{mcauley2015image}\footnote{http://jmcauley.ucsd.edu/data/amazon/}. A dataset that consists of product views from the widely used Amazon platform. We choose the largest subset Book with various types of books for evaluation. The maximum sequence length is set to 20.
    \item \textbf{Gowalla}~\cite{cho2011friendship}. 
    A typical checking-in dataset built from a location-based social networking website. The maximum sequence length is set to 40.
    \item \textbf{RetailRocket}\footnote{https://www.kaggle.com/retailrocket/ecommerce-dataset}. An E-Commerce dataset that contains multiple types of user behaviors over a 4-month period. We only use the view events. The maximum sequence length is set to 20.
\end{itemize}
The dataset preprocessing follows the prior study~\cite{cen2020controllable}. We remove all items and users that occur less than five times in these datasets. All the interactions are regarded as implicit feedback. The statistics of the three datasets after preprocessing are summarized in Table~\ref{tab:match_stats}.

\subsubsection{Training and Evaluation Setup.}

We follow the prior studies~\cite{cen2020controllable,Chen2021PIMI} 
to partition the training, validation, and test sets with a ratio of 8:1:1 regarding unique users
and train models using the entire sequence of the users in the training set. For evaluation, we use the first 80\% of the user behavior sequence to infer the user embeddings and compute the metrics with the remaining 20\% items in the sequence. More details can be found in~\cite{cen2020controllable,Chen2021PIMI}. We adopt the widely used metrics, i.e., \textbf{Recall}, \textbf{Hit Rate}, and \textbf{NDCG} (Normalized Discounted Cumulative Gain)\footnote{The calculation of NDCG follows the officially revised code of ComiRec: https://github.com/THUDM/ComiRec, which is in line with the paper's description of the calculation of NDCG but does not align with the results in the original paper.}, to evaluate our proposed solution. The metrics are computed with the top 20/50 matched candidates.
\subsubsection{Baseline Models.}
We use two types of baselines for performance comparison: general candidate matching and state-of-the-art multi-interest models. 
The baselines are introduced as follows:
\begin{itemize}
    \item \textbf{Most Popular.} A basic algorithm that recommends the most popular items to users. 

    \item \textbf{GRU4Rec}~\cite{gru}. An RNN-based model that captures sequential patterns.
    \item \textbf{YouTube DNN}~\cite{covington2016deep}. A two-tower DNN model that pools behavior embeddings followed by MLP layers to get the final user representation.
    \item \textbf{MIND}~\cite{li2019multi}. The first multi-interest framework that captures diverse interests with capsule networks. 
    \item \textbf{ComiRec-SA}~\cite{cen2020controllable} An advanced multi-interest framework that allows diversity control and further introduces multi-head attention to model multiple interests. We use the ComiRec-SA version, which demonstrates comparable performance with the DR version with a stabler and faster training.
    \item \textbf{Re4}~\cite{zhang2022re4}. An advanced multi-interest framework that takes the backward flow into account to regularize the process. Note that the sampling size is set same as other methods to ensure fairness.
    \item \textbf{$\text{UMI}_\text{HN}$}~\cite{chai2022user}. 
    A variant of UMI. UMI proposes to use user profiles when generating users' multiple interests and proposes an HN strategy to enhance training. Since the HN is shown to be the most effective part of the method and the user profile is not available in the datasets, we use \textbf{$\text{UMI}_\text{HN}$} as our baseline.
    \item \textbf{PIMIRec}~\cite{Chen2021PIMI}. A model that considers time information and interactivity among items when modeling user interests.

\end{itemize}

\subsubsection{Implementation Details.}
We implement our work and baseline methods with PyTorch 1.8 in Python 3.7. We build REMI on \textbf{ComiRec-SA} by default for the best efficiency, except for Section~\ref{Exp-rq2}.
All parameters are set as follows if not otherwise noted:
following \cite{cen2020controllable},
the number of dimensions $d$ for embeddings is set to 64, the batch size is set to 128, and the maximum number of training iterations is set to 1 million for all the models. 
As discussed in Section~\ref{sampled-sm}, the batch-wise shared negative training sample size is set to 128 * 10 as prior works for fair comparison in the main study. 
It is worth mentioning that REMI can get comparable performance with a smaller sample size, which further enables efficient training in practice. We discuss this in Section~\ref{exp-rq3}. 
In the overall comparisons (Table~\ref{tab:match_results},~\ref{tab:improvement}), we report the best performance with the interest number $K \in \{2, 4, 6, 8\}$ for baseline multi-interest models and $K=4$ for REMI. 
In other analysis studies, $K=4$ is used by default.
We use the Adam optimzer~\cite{kingma2014adam} for training. 
For MIND and Re4, we grid search the optimal learning rates from $\{1e-3, 3e-3, 5e-3\}$ and weight decay from $\{1e-6, 1e-5\}$. 
We use $lr = 1e-3$ for REMI and other baselines. 
Other hyperparameters for the baselines are set according to the original paper.
For the hyperparameters for REMI, we search $\beta$ in $\{0.1, 1, 4, 10\}$ for different conditions and set $\lambda$ to $1e2$.

\begin{table*}[t]
    \centering
    \caption{Performance comparison of different methods on three datasets. The bold and underlined numbers represent the best and
second-best results, respectively. The last column shows the relative improvement of REMI over the best baseline methods. We conduct paired t-test of REMI and the best baseline, and the improvement of REMI is significant with $p \leq 0.01$ for all the settings.}
    \resizebox{\textwidth}{!}{
    \begin{tabular}{c c m{1.5cm}<{\centering} m{1.5cm}<{\centering} m{1.5cm}<{\centering} m{1.5cm}<{\centering} m{1.5cm}<{\centering} m{1.5cm}<{\centering} m{1.5cm}<{\centering} m{1.5cm}<{\centering} | m{1.5cm}<{\centering} m{1.5cm}<{\centering}}
    \hline
    \hline
    \multirow{1}*{\textbf{Dataset}} & \multirow{1}*{\textbf{Metric}} & \multicolumn{1}{c}{Pop} & \multicolumn{1}{c}{GRU4Rec} & \multicolumn{1}{c}{Youtube DNN} & \multicolumn{1}{c}{MIND} & \multicolumn{1}{c}{ComiRec} & \multicolumn{1}{c}{Re4} & \multicolumn{1}{c}{$\text{UMI}_{\text{HN}}$}& \multicolumn{1}{c|}{PIMIRec} & \multicolumn{1}{c}{REMI} & \multicolumn{1}{c}{\textbf{Improv.}} \\
    \hline
     \multirow{6}*{Amazon Books} 
              ~ & R@20  &  0.0158 &  0.0441 &  0.0467 &  0.0420 &  0.0557 & 0.0597 & \underline{0.0690} & 0.0682 & \textbf{0.0826} & +19.7\% \\ 
              ~ & HR@20 &  0.0345 &  0.1004 &  0.1043 &  0.0986 &  0.1142 & 0.1240 & \underline{0.1423} & 0.1411 & \textbf{0.1650} & +16.0\% \\
              ~ & ND@20 &  0.0143 &  0.0378 &  0.0391 &  0.0357 &  0.0446 & 0.0476 & \underline{0.0527} & 0.0526 & \textbf{0.0623} & +18.2\% \\
              ~ & R@50  &  0.0281 &  0.0706 &  0.0722 &  0.0687 &  0.0863 & 0.0690 & 0.1053 & \underline{0.1056} & \textbf{0.1189} & +12.6\% \\
              ~ & HR@50 &  0.0602 &  0.1553 &  0.1607 &  0.1533 &  0.1796 & 0.1975 & 0.2059 & \underline{0.2062} & \textbf{0.2298} & +11.4\% \\
              ~ & ND@50 &  0.0193 &  0.0443 &  0.0457 &  0.0433 &  0.0511 & 0.0576 & \underline{0.0587} & 0.0583 & \textbf{0.0657} & +11.9\% \\
              \midrule 

    \multirow{6}*{Gowalla}
              ~ & R@20  &  0.0231 &  0.0900 &  0.0864 &  0.0901 &  0.0805 & 0.0843 & 0.0961 & \underline{0.1193} & \textbf{0.1317} & +10.4\% \\
              ~ & HR@20 &  0.1121 &  0.3359 &  0.3211 &  0.3129 &  0.2901 & 0.3104 & 0.3314 & \underline{0.3843} & \textbf{0.4145} & +7.9\% \\
              ~ & ND@20 &  0.0483 &  0.1433 &  0.1384 &  0.1331 &  0.1210 & 0.1287 & 0.1391 & \underline{0.1603} & \textbf{0.1774} & +10.7\%\\
              ~ & R@50  &  0.0365 &  0.1458 &  0.1388 &  0.1456 &  0.1320 & 0.1396 & 0.1642 & \underline{0.1951} & \textbf{0.2062} & +5.7\% \\
              ~ & HR@50 &  0.1582 &  0.4577 &  0.4390 &  0.4442 &  0.4086 & 0.4224 & 0.4719 & \underline{0.5207} & \textbf{0.5455} & +4.8\% \\
              ~ & ND@50 &  0.0569 &  0.1494 &  0.1434 &  0.1424 &  0.1310 & 0.1410 & 0.1505 & \underline{0.1660} & \textbf{0.1785} & +7.5\% \\
          
         \midrule
            \multirow{6}*{Retail Rocket}
              ~ & R@20  &  0.0129 &  0.0827 &  0.1050 &  0.1171 &  0.1304 & 0.1397 & 0.1519 & \underline{0.1828} & \textbf{0.2109} & +15.3\% \\
              ~ & HR@20 &  0.0252 &  0.1376 &  0.1711 &  0.1883 &  0.1904 & 0.2103 & 0.2364 & \underline{0.2764} & \textbf{0.3144} & +13.7\%\\
              ~ & ND@20 &  0.0098 &  0.0517 &  0.0641 &  0.0698 &  0.0689 & 0.0785 & 0.0875 & \underline{0.1025} & \textbf{0.1170} & +14.1\% \\
              ~ & R@50  &  0.0244 &  0.1371 &  0.1608 &  0.1899 &  0.1922 & 0.2194 & 0.2423 & \underline{0.2811} & \textbf{0.3214} & +14.3\% \\
              ~ & HR@50 &  0.0462 &  0.2132 &  0.2518 &  0.2927 &  0.2895 & 0.3174 & 0.3574 & \underline{0.3969} & \textbf{0.4544} & +14.5\%\\
              ~ & ND@50 &  0.0139 &  0.0593 &  0.0701 &  0.0795 &  0.0786 & 0.0884 & 0.0974 & \underline{0.1093} & \textbf{0.1274} & +16.6\%\\
     \hline \hline
    \end{tabular}       }

    \label{tab:match_results}
\end{table*}

\subsection{Overall Performance (RQ1)}
\label{exp-overall}

We summarize the overall performance in Table~\ref{tab:match_results}. 
Based on the result, we can make the following observations:

First, the personalized deep learning-based models perform much better than MostPopular, indicating the importance of modeling user history behavior in RS. In addition, vanilla multi-interest methods (i.e., MIND and ComiRec) do not always outperform the general learning models like YouTube DNN and GRU4Rec. This is in line with some prior studies on candidate matching~\cite{chai2022user}, which may be attributed to the unsuitable training scheme. 

Also, the enhanced multi-interest models (Re4 and UMI) help to improve the multi-interest model compared with basic ComiRec and MIND, demonstrating the effectiveness of including backward flow and better training procedures. Note that the advanced training strategy in UMI, despite a similar name to our IHN strategy, is from a different aspect. We present a more detailed analysis and compatibility experiments in Section~\ref{Exp-rq2}. We also observe that PIMIRec has the best performance among the baselines, in which the interactivity module demonstrates the most significant performance improvement~\cite{Chen2021PIMI}. It also, to some extent, alleviates the collapse problem through the early interactions among different items. Moreover, in our experiment, we notice that Re4 and PIMIRec require a long training time due to the utilization of complex backward flow and graph structure, leading to efficiency concerns in these solutions.

Finally, REMI, when built on the simple ComiRec-SA, significantly outperforms the state-of-the-art multi-interest models and general models on all three datasets regarding all metrics. Through solving two key questions of the multi-interest learning process, the REMI framework remarkably improves the capacity of existing multi-interest models without architecture modification or extra computation burden. These results demonstrate the importance of the identified problems and the effectiveness of our solutions.

\subsection{Enhancement Study (RQ2)}
\label{Exp-rq2}
\begin{table}[t]
    \centering
    \caption{Generalization experiments. Performance of existing multi-interest models and their REMI-enhanced version.}
    \resizebox{\linewidth}{!}{
    \begin{tabular}{c|c|c c c|c c c|c c c}
    \hline
    \hline
    \multirow{3}*{\textbf{Dataset}} & \multirow{3}*{\textbf{Metric}} & \multicolumn{3}{c|}{ComiRec} & \multicolumn{3}{c|}{PIMIRec} & \multicolumn{3}{c}{$\text{UMI}_{\text{HN}}$} \\
              ~ & & Original  & +REMI & +$\Delta$ & Original  & +REMI & +$\Delta$ & Original  & +REMI  & +$\Delta$ \\
    \hline
     \multirow{6}*{Books} 
              ~ & R@20 &  0.0557 &  \textbf{0.0826} & +48.29\% &  0.0682 &  \textbf{0.0834} & +22.28\% &  0.0690 &  \textbf{0.0828} & +20.00\% \\
              ~ & HR@20 &  0.1142 &  \textbf{0.1650} & +44.48\% &  0.1411 &  \textbf{0.1678} & +18.92\% &  0.1423 &  \textbf{0.1672} & +17.49\%  \\
              ~ & ND@20 &  0.0446 &  \textbf{0.0623} & +39.68\% &  0.0526  & \textbf{0.0643} & +22.24\% &  0.0527 &  \textbf{0.0637} & +20.87\%\\
              ~ & R@50 &  0.0863 &  \textbf{0.1189} & +37.77\%&  0.1056 &  \textbf{0.1204} & +14.01\%&  0.1053 &  \textbf{0.1206} & +14.52\% \\
              ~ & HR@50 &  0.1796 &  \textbf{0.2298} & +27.95\% &  0.2062 &  \textbf{0.2312} & +12.12\% &  0.2059 &  \textbf{0.2335} & +13.40\%\\
              ~ & ND@50 &  0.0511 &  \textbf{0.0657} & +28.57\% &  0.0583 &  \textbf{0.0673} & +15.43\% & 0.0587 & \textbf{0.0675} & +14.99\% \\
              \midrule 

    \multirow{6}*{Gowalla}
              ~ & R@20 &  0.0805 &  \textbf{0.1317} & +63.60\% & 0.1193 &  \textbf{0.1327} & +11.23\% & 0.0961 &  \textbf{0.1328} & +38.18\% \\
              ~ & HR@20 &  0.2901 &  \textbf{0.4145} & +42.88\% &  0.3843 &  \textbf{0.4231} & +10.09\% &  0.3314 & \textbf{0.4194} & +26.55\%\\
              ~ & ND@20 &  0.1210 &  \textbf{0.1774} & +46.61\% &  0.1603 &  \textbf{0.1785} & +11.35\% &  0.1391 &  \textbf{0.1772} & +27.39\% \\
            ~ & R@50 &  0.1320 &  \textbf{0.2062} & +56.21\%&  0.1951 &  \textbf{0.2143} & +9.84\% &  0.1642 &  \textbf{0.2102} & +28.01\%  \\
              ~ & HR@50 &  0.4086 &  \textbf{0.5455} &+33.50\% &  0.5207 &  \textbf{0.5632} & +8.16\% &  0.4719 &  \textbf{0.5571} & +18.05\% \\
              ~ & ND@50 &  0.1310 &  \textbf{0.1785} & +36.25\% &  0.1660 &  \textbf{0.1826} & +10.00\% & 0.1505 &  \textbf{0.1816} & +20.66\% \\
          
         \midrule
            \multirow{6}*{\makecell[c]{Retail \\ Rocket}}
              ~ & R@20 &  0.1304 &  \textbf{0.2109} & +61.73\% &  0.1828 &  \textbf{0.2173} & +18.87\% &  0.1519 &  \textbf{0.2155} & +41.86\% \\
              ~ & HR@20 &  0.1904 &  \textbf{0.3144} & +65.12\% & 0.2764 &  \textbf{0.3265} & +18.12\% & 0.2364 &   \textbf{0.3236} & +36.88\% \\
              ~ & ND@20 &  0.0689&  \textbf{0.1170} & +69.81\% &  0.1025 &  \textbf{0.1210} & +18.04\% & 0.0875 & \textbf{0.1211} & +38.40\% \\
              ~ & R@50 &  0.1922 &  \textbf{0.3214} & +67.22\% & 0.2811 &  \textbf{0.3217} & +14.43\% &  0.2423 &  \textbf{0.3161} & +30.45\% \\
              ~ & HR@50 &  0.2895 &  \textbf{0.4544} & +56.96\%& 0.3969 &  \textbf{0.4587} & +15.57\% &  0.3574 & \textbf{0.4464} & +24.90\%\\
              ~ & ND@50 &  0.0786 &  \textbf{0.1274} & +62.08\% &  0.1093 &  \textbf{0.1298} & +18.75\% &  0.0974  &  \textbf{0.1259} & +29.26\%\\

     \hline \hline
     
    \end{tabular}
}

    \label{tab:improvement}
\end{table}

As a general training framework with simple implementation, REMI can be applied to the existing multi-interest models to further boost their performance. In this section, we include a comparison of the original ComiRec-SA~\cite{cen2020controllable}, PIMIRec~\cite{Chen2021PIMI}, $\text{UMI}_\text{HN}$~\cite{chai2022user} with their REMI-enhanced versions. The results are shown in Table~\ref{tab:improvement}.

ComiRec represents the basic multi-interest framework with no additional information or enhancement. 
We observe improvements from \textbf{27.95\% to 69.81\%} on this simple model on the three datasets.
 These improvements reflect the importance of the identified problems and the effectiveness of the proposed solutions in basic multi-interest learning.
They also suggest the potential for the application of REMI in large-scale industry datasets since it helps achieve strong performance even with simple and light architecture.
PIMIRec represents an advanced multi-interest model that integrates additional information (i.e., periodicity) and incorporates complex architecture such as GNN  to tackle the evolutional user interests. 
By explicitly regularizing the routing process and incorporating effective IHN training techniques, our solution can further enhance the performance of PIMI. 
We also combine REMI with $\text{UMI}_\text{HN}$.
$\text{UMI}_\text{HN}$ improves the training by applying a candidate-aware calculation of scores instead of selecting one interest for all candidates as \autoref{eqn:argmax}. However, it still attaches the same importance to the uniformly-sampled negative items and also leads to lower efficiency, while the IHN in REMI approximates an interest-aware hard negative sampling with negligible cost. The results show that REMI is compatible with their optimization and leads to further performance improvement.

\subsection{Effectiveness of Interest-aware Hard Negative Mining (RQ3)}
\label{exp-rq3}

In this section, we first present a quantitative analysis of the efficiency and effectiveness of the proposed IHN (Interest-aware Hard Negative Mining) method for multi-interest model training. 
Afterward, we present several case studies to substantiate whether the IHN method assigns higher weights to those interest-aware hard negatives.

\begin{figure}[t]

    \begin{subfigure}{\linewidth}
        \centering
\includegraphics[width=1.0\linewidth]{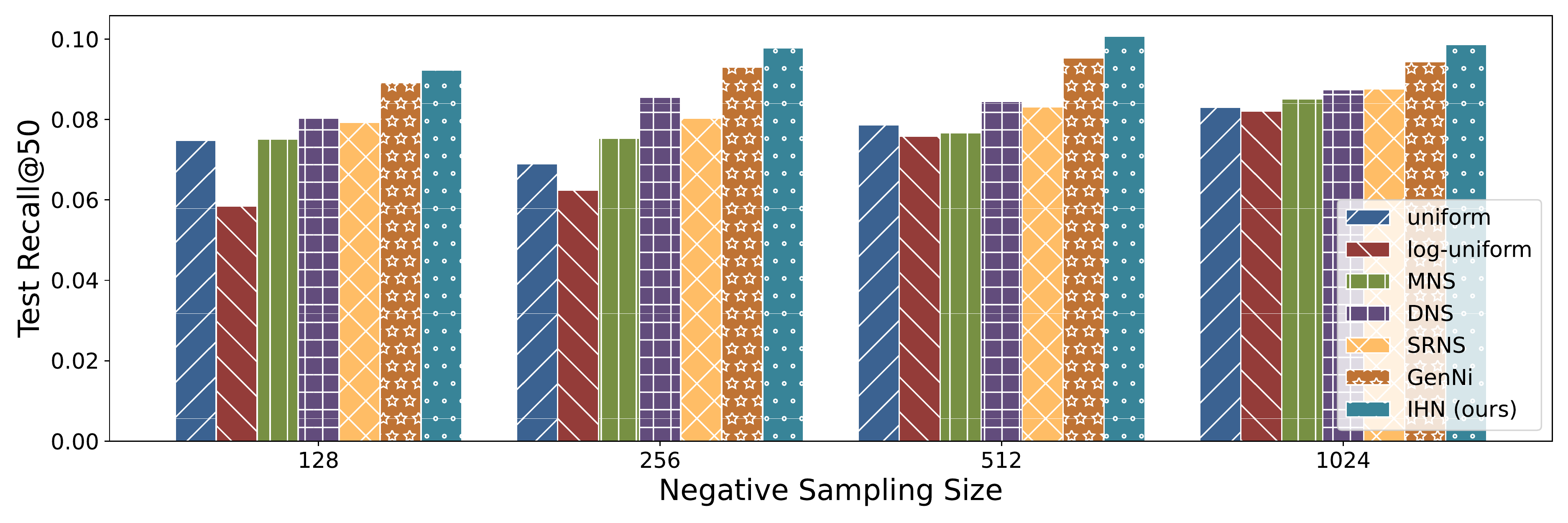}
    \end{subfigure}

  \caption{The effectiveness  comparison of different negative sampling strategies on the Amazon Book dataset in Recall@50.}   
   \label{fig:exp-valid-training}
\end{figure}

\begin{figure}[t]

    \begin{subfigure}{0.49\linewidth}
        \centering
        \includegraphics[width=1.0\linewidth]{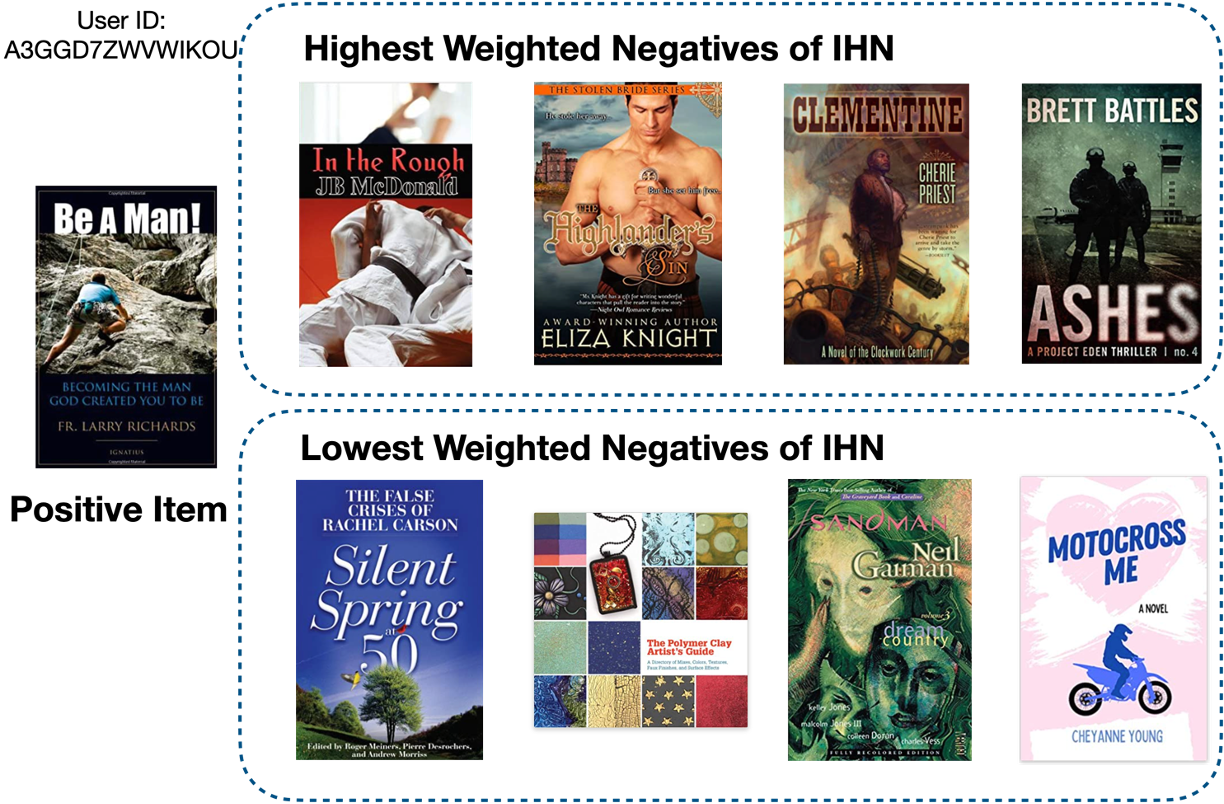}
    \end{subfigure}
    \begin{subfigure}{0.495\linewidth}
        \centering
        \includegraphics[width=1.0\linewidth]{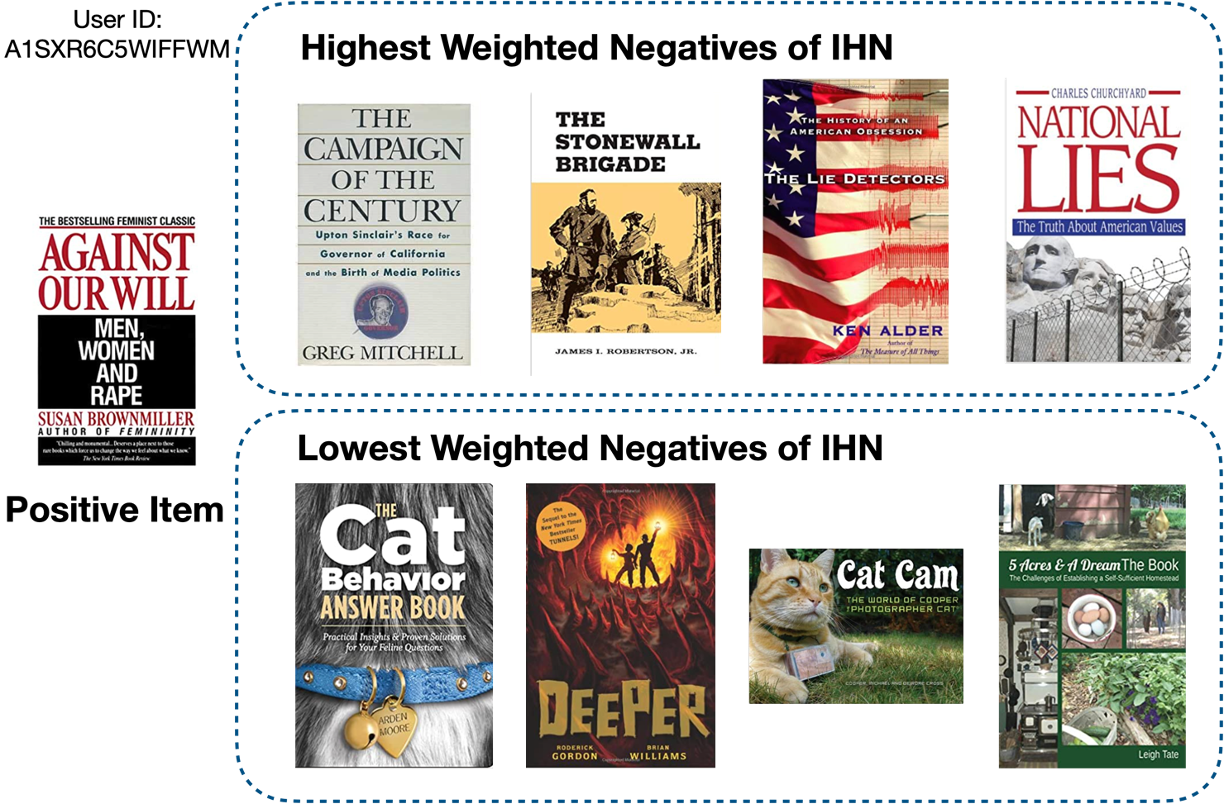}
    \end{subfigure}

  \caption{Case Study on Amazon Book. IHN assigns higher weights to the hard negatives according to the current interest, thus making the learning of user interest representations more discriminative. }   
   \label{fig:case}
\end{figure}

For a fair comparison, we do not include routing regularization and compare IHN to the generic uniform sampling and log-uniform sampling, as well as the state-of-the-art sampling techniques, i.e., MNS~\cite{yang2020mixed}, DNS~\cite{zhang2013optimizing}, SRNS~\cite{ding2020simplify}, and GenNi~\cite{chen2022generating}, when applied to ComiRec-SA.
We observe similar results for three datasets, and due to space limitations, we present the results on the Amazon Book dataset.
IHN consistently outperforms all baselines in multi-interest learning, as demonstrated in \autoref{fig:exp-valid-training} with efficiency similar to uniform sampling. 
For sampling techniques that exhibit similar efficiency, such as uniform and log-uniform sampling, we observe that the performance gap between IHN and generic sampling increases as the negative sampling size decreases, indicating the potential of IHN in further improving training efficiency with a smaller negative sampling size without compromising performance.
In addition, the commonly adopted log-uniform loss is found to be inferior to uniform sampling when dealing with large item candidate pools. This is due to the fact that the log-uniform distribution assigns significantly small sampling probabilities to unpopular items in such cases. Consequently, this exacerbates the popularity bias and leads to undertrained embeddings for unpopular items.

Next, we present case studies to validate whether \textbf{IHN indeed assigns higher weights to interest-aware hard negatives}. 
We randomly select several users from the Amazon Book Dataset. 
Since user interest representations are implicit and can not be directly visualized, we display the true positive items that can be used to imply the selected interest concept in the multi-interest learning framework, as demonstrated in \autoref{fig:overview}.
We present the negative items with the highest and lowest weights assigned by the IHN method in \autoref{fig:case}.
The left figure shows that when the positive sample relates to manhood and faith, the highest-weighted items are hard negatives related to the heroic spirit, while the lowest-weighted items are entirely unrelated books, such as those about art, which do not provide useful information for training discriminative user embeddings.
Similarly, in the right figure, when the positive sample pertains to politics and humanity, particularly feminism, the highest-weighted negative samples of IHN are also related to politics but not feminism, providing more granular supervision signals. In contrast, the lowest-weighted negative samples are unrelated, such as those about cats. 
These two examples demonstrate that the IHN method effectively assigns higher weights to more meaningful hard negatives that help train discriminative user interest representation while giving lower weights to entirely unrelated negative samples.

\subsection{Effectiveness of Routing Regularization (RQ4)}

\begin{figure}[t]
    \begin{subfigure}{0.24\linewidth}
        \centering
        \includegraphics[width=1.0\linewidth]{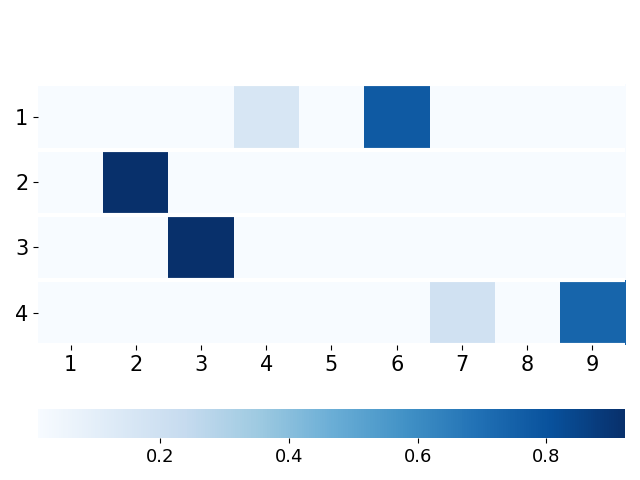}
        \subcaption{PIMIRec}
    \end{subfigure}
    \begin{subfigure}{0.24\linewidth}
        \centering
        \includegraphics[width=1.0\linewidth]{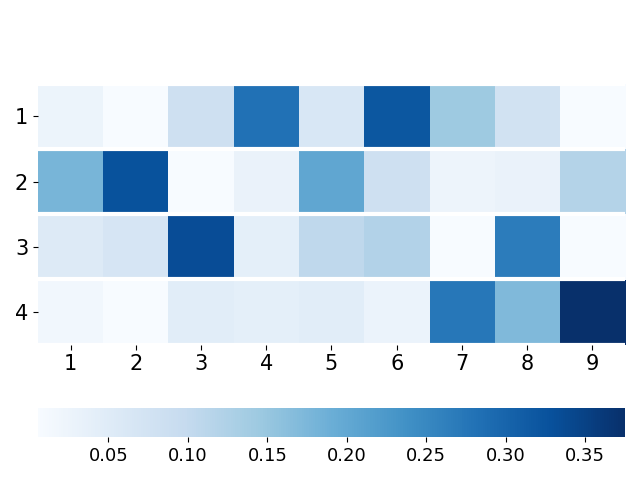}
        \subcaption{REMI+PIMI} 
    \end{subfigure}
        \begin{subfigure}{0.24\linewidth}
        \centering
        \includegraphics[width=1.0\linewidth]{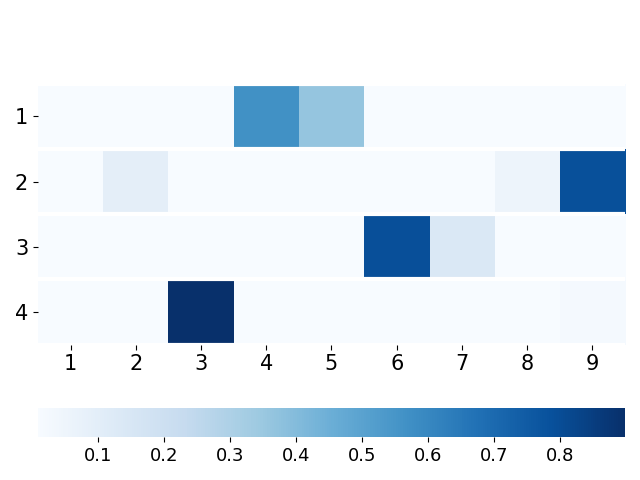}
        \subcaption{UMI}
    \end{subfigure}
    \begin{subfigure}{0.24\linewidth}
        \centering
        \includegraphics[width=1.0\linewidth]{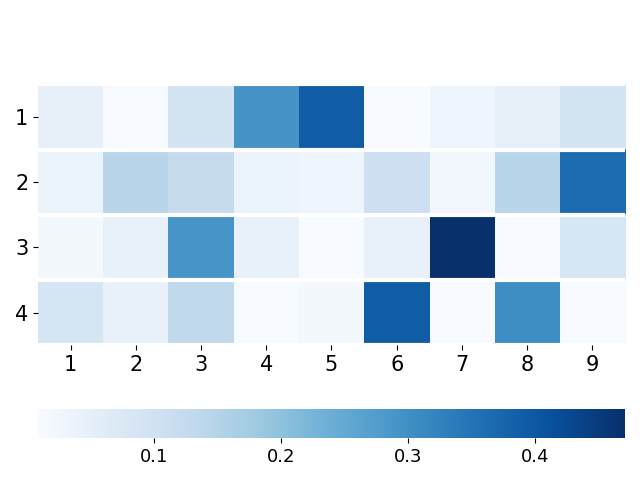}
        \subcaption{REMI+UMI} 
    \end{subfigure}
    \caption{Visualization of examples of routing matrices for PIMIRec, UMI, and their REMI-enhanced versions. REMI solves the routing collapse problem.}
    \label{fig:vis_more}
\end{figure}

This section verifies the effectiveness of Routing Regularization without adding IHN to ensure fairness with other methods. We compared it to other multi-interest regularization methods, including Re4~\cite{zhang2022re4} and SINE~\cite{tan2021sparse}.
SINE regulates multi-concepts in the representation space, while Re4 uses backward flow to regulate the training process. 
However, these methods fail to address the \textit{routing collapse} issue and perform worse than our Routing Regularization, as shown in \autoref{tab:rr}. Furthermore, besides the visualization in the introduction, Figure~\ref{fig:intro_sample} (b), we also visualize the item-to-interest routing for state-of-the-art UMI~\cite{chai2022user} and PIMI~\cite{Chen2021PIMI} in Figure~\ref{fig:vis_more}, with and without REMI. Some random samples show that REMI effectively avoids \textit{routing collapse} for UMI and PIMI, where interests are no longer composed of one or two items but fuse all corresponding relevant items from the historical sequence.

\begin{table}[t]
\caption{Comparison of various regularization strategies. We show the metrics@50 on three datasets. The final column in each dataset block displays the HR@50 improvement. }
\label{tab:rr}
\resizebox{\linewidth}{!}{
\centering
\begin{tabular}{c|cccc|cccc|cccc}
    \hline \hline
  & \multicolumn{4}{c|}{\textbf{Books}}& \multicolumn{4}{c|}{\textbf{Gowalla}}& \multicolumn{4}{c}{\textbf{RetailRocket}} \\
\textbf{Method}  &  Recall &  NDCG & HR & +$\Delta$& Recall &  NDCG & HR & +$\Delta$& Recall &  NDCG & HR & +$\Delta$
\\ 
    \hline
No Reg.  & 0.0824 & 0.0464 & 0.1661 & - & 0.1271 & 0.1246 & 0.3891 & - & 0.1818 & 0.0761 & 0.2738 & -\\ 
Rep. Reg.~\cite{tan2021sparse} & 0.0814 & 0.0481 & 0.1712 & +3.07\% & 0.1141 & 0.1261 & 0.3935 & +1.13\% & 0.1895 & 0.0822 & 0.2939 & +7.34\%\\ 
Re4 & 0.1008 & 0.0562 & 0.1975 & +18.90\% & 0.1182 & 0.1309 & 0.4060 & +4.34\% & 0.2088 & 0.0884 & 0.3174 & +15.92\%\\ 
RR (Ours) & 0.1078 & 0.0606 & 0.2135 & +28.53\% & 0.1841 & 0.1626 & 0.5067 & +30.22\% & 0.2804 & 0.1145 & 0.4028 & +47.11\%\\ 
    \hline \hline
\end{tabular}}
\end{table}
\subsection{Ablation Study and Hyper-parameter Study (RQ5)}
\label{exp-rq4}
\begin{table}[t]
\caption{Ablation study of Interest-aware Hard Negative Sampling (IHN) and Routing Regularization(RR). We show the metrics@50 on three datasets. The final column in each dataset block displays the HR@50 improvement.}
\resizebox{\linewidth}{!}{
\centering
\begin{tabular}{cc|cccc|cccc|cccc}
    \hline \hline
 \multicolumn{2}{c|}{\textbf{Settings}} & \multicolumn{4}{c|}{\textbf{Books}}& \multicolumn{4}{c|}{\textbf{Gowalla}}& \multicolumn{4}{c}{\textbf{RetailRocket}} \\

\textsc{IHN}  & \textsc{RR}  & Recall & NDCG & HR & +$\Delta$  & Recall & NDCG & HR & +$\Delta$  & Recall & NDCG & HR & +$\Delta$
\\ 
    \hline
\xmark& \xmark & 0.0814 & 0.0481 & 0.1661 & - & 0.1271 & 0.1246 &  0.3891 & - & 0.1818 & 0.0761 & 0.2738 & -\\ 
\cmark& \xmark & 0.1008 & 0.0562&  0.1975 & +18.90\% & 0.1608 & 0.1525 & 0.4715 & +21.17\% & 0.2559 & 0.0907 &0.3307 & +20.78\%\\  
\xmark& \cmark & 0.1078 & 0.0606 & 0.2135 & +28.53\% & 0.1841 & 0.1626 &  0.5067 & +30.22\% & 0.2804 & 0.1145 & 0.4028 & +47.11\%\\ 
\cmark& \cmark & 0.1189 & 0.0657 & 0.2298 & +38.35\% & 0.2060 & 0.1783 & 0.5455 & +40.19\% & 0.3124 & 0.1274 & 0.4544 & +65.96\%\\ 
    \hline \hline
\label{tab:ablation}
\end{tabular}}
\end{table}
\subsubsection{Ablation Studies}
Here we present a detailed analysis of the effect of Interest-aware Hard Negative Mining (IHN) and Routing Regularization (RR) (Table~\ref{tab:ablation}).
Through solving two key problems in Multi-Interest Recommendation, IHN and RR independently improve performance remarkably. Combined, they boost the performance with a \textbf{38.35\% to 65.96\%} improvement in Hit Rate@50 on the three datasets. Noteworthy, these improvements come with negligible costs of computation and latency because the two proposed strategies only modify the training objective of the original ComiRec-SA model by re-weighting the loss and adding a variance-based regularization term.


\subsubsection{Hyperparameters Studies} 
\label{exp-rq4-hyper}
In this section, we thoroughly study three hyperparameters in REMI: the interest number $K$, concentration parameter $\beta$, and balance parameter $\lambda$.

\begin{table}[]
\centering
\caption{Study on the interest number K. We show the metrics@50 on three datasets.}
\label{tab:interest-k}
\resizebox{0.9\columnwidth}{!}{%
\begin{tabular}{c|ccc|ccc|ccc}
\hline
\hline
 & \multicolumn{3}{c|}{\textbf{Books}} & \multicolumn{3}{c|}{\textbf{Gowalla}} & \multicolumn{3}{c}{\textbf{RetailRocket}} \\
K & HR & Recall & NDCG & HR & Recall & NDCG & HR & Recall & NDCG \\ \hline
2 & 0.2182 & 0.1129 & 0.0625 & 0.5503 & 0.2059 &0.1786 & 0.4461 & 0.3181& 0.1241 \\
4 & 0.2298 & 0.1189 & 0.0657 & 0.5455 & 0.2062 & 0.1785 & \underline{0.4544} & \textbf{0.3214}& \textbf{0.1274} \\
6 & \underline{0.2340} & \underline{0.1216} & \underline{0.0667} & \underline{0.5515} & \underline{0.2065} & \underline{0.1790} & \textbf{0.4550} & \underline{0.3210}& \underline{0.1267} \\
8 & \textbf{0.2360} & \textbf{0.1220} & \textbf{0.0673} & \textbf{0.5565} & \textbf{0.2106} & \textbf{0.1793} & 0.4470 & 0.3193& 0.1254 \\ \hline \hline
\end{tabular}%
}
\end{table}
\textbf{The interest number $K$.} First, we show the performance of different interest numbers K of REMI in Table~\ref{tab:interest-k}.
In the study,  we fix the concentration parameter $\beta$ and  balance parameter $\lambda$ to be 1 and 1e2, respectively.  
Generally, REMI demonstrates better performance with a larger $K$, indicating that REMI can leverage the strength of multi-interest encoding and generate more representative and diverse interests when the interest number grows.
Furthermore, the optimal number of interests also varies depending on the dataset. For instance, in the case of RetailRocket, where the average user sequence length and the item corpus are relatively small, four interests results in the best performance.

\begin{figure*}[t]
    \begin{subfigure}{0.245\linewidth}
        \centering
        \includegraphics[width=1.0\linewidth]{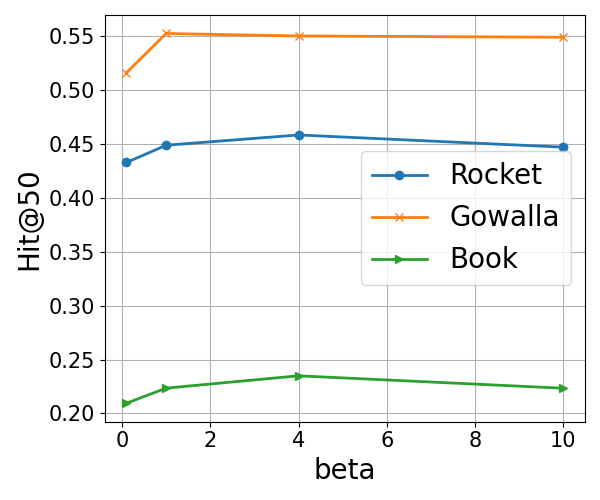}
    \end{subfigure}
    \begin{subfigure}{0.245\linewidth}
        \centering
        \includegraphics[width=1.0\linewidth]{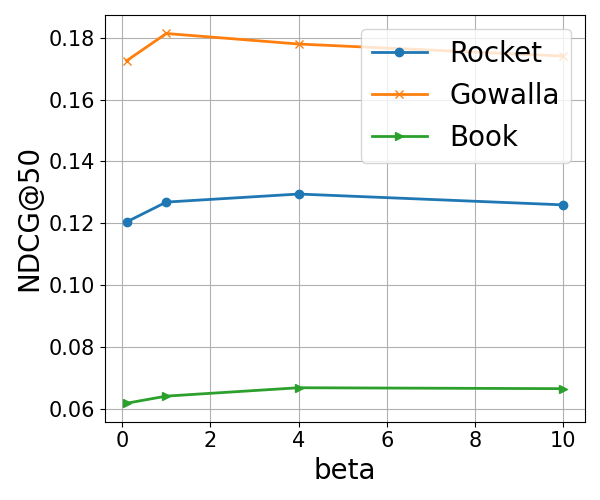}
    \end{subfigure}
    \begin{subfigure}{0.245\linewidth}
        \centering
        \includegraphics[width=1.0\linewidth]{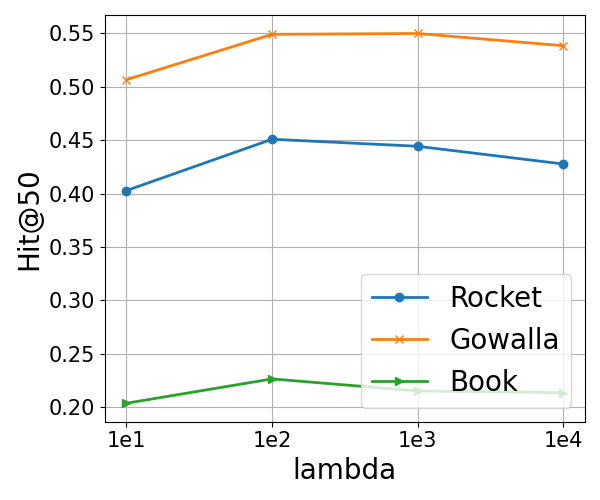}
    \end{subfigure}
    \begin{subfigure}{0.245\linewidth}
        \centering
        \includegraphics[width=1.0\linewidth]{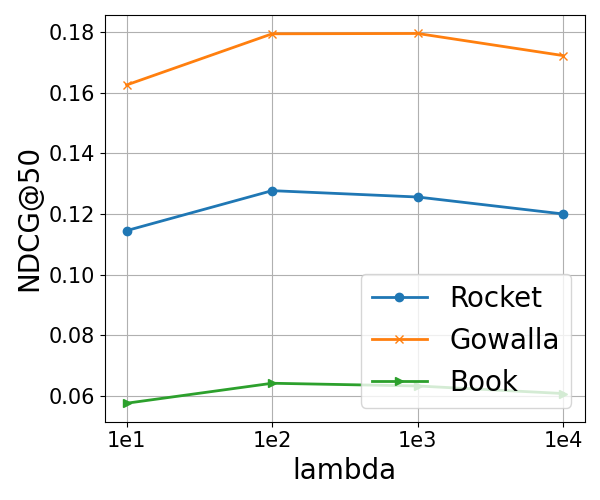}
    \end{subfigure}
    
    \caption{Study on balance parameter $\beta$ and $\lambda$. We show Hit Rate@50 and NDCG@50 on three datasets.}
    \label{fig:exp-lambda}
\end{figure*}
\textbf{The concentration parameter $\beta$.} 
We investigate $\beta$ in the range of $0.1$ to $10$ to enable its exponential coverage to cover a majority of values, as shown in Figure~\ref{fig:exp-lambda}.
In the study, we fix the balance parameter $\lambda$ and interest number $K$ to be 1e2 and 4, respectively.
We observe that larger $\beta$ does not necessarily lead to improvement. 
It aligns with the analysis in Section~\ref{method-ihn} that easy negatives are also important to supervise the training of RS, and the hardest samples might be false negatives. 
Moreover, the optimal choices of $\beta$ for different datasets differ. 
Specifically, for Amazon Book and RetailRocket, the optimal $\beta$ in the set is 4, while setting $\beta$ as 1 produces the best performance for Gowalla. 
This can be attributed to the different characteristics of the dataset, such as the property of items. 
The result also demonstrates the importance of an adjustable $\beta$ under different scenarios.

\textbf{The balance parameter $\lambda$.}  Figure~\ref{fig:exp-lambda} presents the Hit Rate@50 and NDCG@50 of REMI with different $\lambda$. 
In order to balance the loss between two differing magnitudes, we explore the effects of varying the magnitude of $\lambda$.
In the study, we fix the interest number $K$ and concentration parameter $\beta$ to be 4 and 1, respectively.
It shows that as $\lambda$ increases, the performance first increases and then decreases. That aligns with our intuition that we need to properly select $\lambda$ to avoid \textit{routing collapse} while not influencing the training to force the routing to be average. 
We roughly tune it from $\{1e1, 1e2, 1e3, 1e4\}$ and fix $\lambda$ to be $1e2$ for other studies.

\section{Conclusions}
This work revisits the existing training scheme of multi-interest learning and reveals the issues of increased easy negatives and routing collapse. To address these challenges, we propose REMI as a general training framework. REMI first mitigates the problem of easy negatives with an ideal interest-aware hard negative sampling distribution and an approximation method to achieve the goal at a negligible computational cost. REMI also incorporates a novel routing regularization to avoid routing collapse and further improve the modeling capacity of multi-interest models. Extensive experiments demonstrate that training with the REMI framework significantly boosts the performance of existing methods.
We hope this training framework will further contribute to future research on promising multi-interest learning and highlight the importance of exploring training procedures in related fields.
\appendix

\section{Implementation of IHN}
\label{supp-code}
We provide PyTorch-style pseudocode in Algorithm~\ref{algo} for our proposed IHN, compared with the original uniformly sampled softmax.
\definecolor{myblue}{RGB}{0, 90, 146}
\definecolor{myred}{RGB}{160, 0, 10}
\setlength{\belowcaptionskip}{-0.1cm} 
\begin{algorithm}
	\caption{IHN.} 
	\begin{algorithmic}[1]
	    \State {\color{myblue}{\# pos : exp of inner product for the positive item}}
	    \State {\color{myblue} \# neg : exp of inner products for the sampled negative items}
	    \State {\color{myblue}\# beta : concentration parameter}
	    \State
	    \State {\color{myblue}\# objective with uniform negative sampling}
     \State \color{black}{Neg = neg.{\color{myred}{sum}}{\small{$()$}} }
	    \State \color{black}{loss\_uniform = -log{\small{(}}pos / {\small{(}}pos + Neg{\small)) } }
        \State
	    \State {\color{myblue} \# objective with IHN}
	    \State \color{black}{imp = (beta* log{\small{(}}neg{\small{)}}}).{{exp}}{\small{$()$}}
	    \State \color{black}{Neg = (imp*neg).{\color{myred}{sum}}{\small{$()$}} / imp.{\color{myred}{mean}}{\small{$()$}}}
        \State loss\_IHN = -log{\small{(}}pos / {\small{(}}pos + Neg{\small)) } 
	\end{algorithmic} 
 \label{algo}
\end{algorithm}


\bibliographystyle{ACM-Reference-Format}
\bibliography{sample-base.bib}

\end{document}